\documentclass[cameraready]{Interspeech}

\usepackage{tabularx}
\usepackage{booktabs}
\usepackage{siunitx}
\usepackage{multirow}
\usepackage{xcolor}
\usepackage{comment}

\title{Do EEG Foundation Models Transfer to Speech? A Benchmark on Overt and Imagined Speech Decoding}

\author[affiliation={1,2}, orcid=0000-0002-0961-0046]{Owais Mujtaba}{Khanday}
\author[affiliation={1,2}, orcid=0009-0003-1440-8314]{Mohamed Baha}{Ben Ticha}
\author[affiliation={3}, orcid=0009-0007-0740-2511]{Sanae}{Belfrouh}
\author[affiliation={4}, orcid=0000-0001-5117-1094]{Marc}{Ouellet}
\author[affiliation={1,2}, orcid=0000-0002-5531-8994]{Jose A.}{Gonzalez-Lopez}

\address{
    $^1$ Dept. Signal Theory, Telematics and Communications, University of Granada, Spain \\
    $^2$ Research Centre for Information and Communication Technologies (CITIC-UGR), Spain \\
    $^3$ Laboratory of Information Technologies, University of Chouaib Doukkali, Morocco \\
    $^4$ Brain, Mind, and Behavior Research Center (CIMCYC), University of Granada, Spain
}

\email{owaismujtaba@ugr.es, joseangl@ugr.es}


\begin{document}

\maketitle

\begin{abstract}
EEG foundation models pretrained on thousands of hours have shown large gains over task-specific networks for motor imagery, seizure detection, sleep staging, and emotion recognition, but their transfer to speech decoding—arguably the most demanding non-invasive BCI application—remains untested. We present the first systematic benchmark of EEG foundation models against strong convolutional baselines for speech decoding, using two corpora: UGR-MINDVOICE (overt and covert Iberian Spanish) and BCI Competition 2020 Track 3 (imagined speech). We compare two foundation models (LaBraM, EEGMamba) against three established baselines (EEGNet, ShallowFBCSPNet, EEGConformer) under a unified preprocessing and fine-tuning protocol. Large-scale EEG pretraining yields no consistent advantage over a 16K-parameter CNN on speech tasks, indicating that current general-purpose EEG pretraining does not yet transfer to speech production and motivating speech-specific foundation models.
\end{abstract}

\section{Introduction}
\label{sec:intro}

Speech impairments caused by brainstem injury or amyotrophic lateral sclerosis (ALS) are debilitating and often lead to social isolation. A speech Brain–Computer Interface (BCI) offers an alternative means of communication in such cases, bridging the brain and the external world by decoding neural signals into an acoustic waveform or text. The most remarkable advances in speech decoding have been achieved with invasive techniques — intracortical arrays \cite{willett2023, card2024, wairagkar_instantaneous_2024, kunz_inner_2025, jude_decoding_2026} and electrocorticography (ECoG) \cite{metzger_generalizable_2022, metzger_high-performance_2023, silva_bilingual_2024, littlejohn_streaming_2025, ticha_neural_nodate, offenberg_optimal_2025} — yet attention is increasingly turning toward less invasive approaches such as stereo-EEG (sEEG) \cite{angrick_real-time_2021, kohler_synthesizing_2022, angrick_online_2024, verwoert_whole-brain_2025, verwoert_moving_2025, singh_transfer_2025}, and toward fully non-invasive ones including magnetoencephalography (MEG) and electroencephalography (EEG) \cite{defossez_decoding_2023, dascoli_towards_2025, li_brainecho_nodate}. EEG is particularly attractive: it is inexpensive, wearable, and safe, requiring no surgical intervention. These advantages, however, come at a cost — low spatial resolution, a poor signal-to-noise ratio, and high sensitivity to ocular and muscular artifacts — which have so far confined EEG speech decoding to small corpora with modest accuracy. Despite these limitations, EEG continues to attract interest, and closing the gap between invasive and non-invasive decoding is widely regarded as one of the field's central challenges.

One reason for optimism comes from an analogy with other domains. In natural language processing and computer vision, scaling self-supervised models to massive unlabeled corpora produced large transferable gains that small task-specific models could not match~\cite{devlin2019}. A similar trajectory may be unfolding for EEG: a new generation of \emph{EEG foundation models} learns generic representations through self-supervised pre-training on thousands of hours of heterogeneous recordings, and is then fine-tuned for downstream tasks. LaBraM~\cite{jiang2024} pre-trains a neural Transformer on roughly 2,500 hours of EEG via vector-quantized neural spectrum prediction, while EEGMamba~\cite{wang2025eegmamba} uses a bidirectional state-space (Mamba) backbone trained on over 16,000 hours. Both report state-of-the-art results across abnormal detection, seizure detection, emotion recognition, motor imagery, and sleep staging, consistently surpassing established convolutional baselines.

Crucially, however, these models have barely been evaluated on \emph{speech and language} tasks. Speech production engages cortical dynamics --- in ventral sensorimotor and superior temporal regions --- that differ markedly from the paradigms (resting state, motor imagery, clinical events) that dominate foundation-model pre-training corpora. It is therefore an open and consequential question whether the representations learned by these models transfer to speech decoding at all, or whether speech requires dedicated pre-training data. The single point of contact in the literature is indirect: EEGMamba reports results on one imagined-speech dataset (BCI Competition 2020 Track~3) as one of six downstream tasks, but without a controlled comparison focused on speech, and with a risk of train/test overlap that we discuss below.

In this work we provide the first systematic answer. We benchmark two recent EEG foundation models (LaBraM, EEGMamba) against three strong baselines (EEGNet, ShallowFBCSPNet, EEGConformer) on overt and imagined speech decoding, using two complementary corpora and a single unified preprocessing and fine-tuning pipeline. Our contributions are:
\begin{itemize}
  \item the first controlled benchmark of EEG foundation models versus convolutional baselines specifically for speech/language decoding;
  \item an evaluation spanning \emph{overt}, \emph{covert}, and \emph{imagined} speech across two datasets (UGR-MINDVOICE; BCIC2020-T3)
\end{itemize}

\section{Methodology}
\label{sec:method}

\subsection{Datasets}
\label{sec:datasets}

We evaluate on two complementary corpora, summarised in Table~\ref{tab:datasets}.
\noindent\textbf{UGR-MINDVOICE}~\cite{ugrmindvoice2025} is a BIDS-compliant corpus of 64-channel EEG (10--20 layout, 1000\,Hz, referenced to FCz) from 15 native Iberian Spanish speakers, with one or two sessions each (22 sessions in total). Within each session, participants produced both \emph{overt} (spoken-aloud) and \emph{covert} (silently imagined) speech, organised in a mini-block design that alternates conditions every 20 trials to limit motor contamination of covert trials. The stimulus set covers all Spanish phonemes (95 CV syllables), 60 words across six semantic categories balanced for lexical frequency, and 30 pseudowords; overt productions are recorded acoustically. 

\noindent\textbf{BCI Competition 2020, Track~3}~\cite{lee2020} contains 64-channel EEG from 15 healthy subjects performing imagined speech of five words/phrases selected for assistive communication --- \emph{hello}, \emph{help me}, \emph{stop}, \emph{thank you}, and \emph{yes}. The dataset provides predefined training, validation and test splits, with the best competition entry reaching 82.6\% on validation and far lower accuracy on the held-out test set~\cite{jeong2022bci}. We include it to (i) add a purely imagined-speech task, (ii) provide an external, widely-used reference point, and (iii) enable comparison with the single previously published foundation-model number on this task~\cite{wang2025eegmamba}.

We verified that BCIC2020-T3 is absent from the pre-training data of both foundation models, ensuring a leakage-free comparison. As the official test labels are not publicly available, we report results on the official validation split.

\begin{table}[t]
  \caption{Benchmark datasets. Ov.=overt, Cov.=covert, Imag.=imagined.}
  \label{tab:datasets}
  \centering
  \scriptsize
  \setlength{\tabcolsep}{4pt}
  \begin{tabularx}{\linewidth}{Xccccc}
    \toprule
    \textbf{Dataset} & \textbf{Lang.} & \textbf{Subj.} & \textbf{Ch.} & \textbf{Mode} & \textbf{Classes} \\
    \midrule
    UGR-MINDVOICE~\cite{ugrmindvoice2025} & ES & 15 & 64 & Ov.+Cov. & 3/6/30 \\
    BCIC2020-T3~\cite{lee2020}            & EN & 15 & 64 & Imag.    & 5 \\
    \bottomrule
  \end{tabularx}
\end{table}

\subsection{Models}
\label{sec:models}

All models receive EEG epoch tensors and produce logits over the task classes through a linear head. We use the braindecode implementations~\cite{schirrmeister2017} for the baselines and the publicly released checkpoints for the foundation models.

\noindent\textbf{Baselines.} \emph{EEGNet}~\cite{lawhern2018} is a depthwise-separable convolutional network: a temporal convolution extracts narrow-band content, a depthwise spatial convolution learns one spatial filter per band, and a separable convolution recombines them, for $\approx$16K parameters. \emph{ShallowFBCSPNet}~\cite{schirrmeister2017} mirrors FBCSP with a temporal convolution, a spatial convolution, a squaring non-linearity, average pooling, and a log transform ($\approx$162K parameters). \emph{EEGConformer}~\cite{song2023} couples a convolutional front-end with a Transformer encoder applying multi-head self-attention over temporal tokens ($\approx$406K parameters).

\noindent\textbf{Foundation models.} \emph{LaBraM}~\cite{jiang2024} segments EEG into channel patches, encodes them with a temporal encoder, and adds learnable temporal and spatial embeddings before a Transformer encoder. It is pre-trained by predicting discrete neural tokens of masked patches, where a vector-quantised tokenizer is trained to reconstruct the Fourier amplitude and phase. We initialise from the public LaBraM-Base checkpoint ($\approx$5.8M parameters; the only publicly released size) and fine-tune end-to-end.
\emph{EEGMamba}~\cite{wang2025eegmamba} reorganises channels into patches, encodes each with a time-domain and an FFT-based frequency-domain module plus conditional positional encoding, and models the sequence with a bidirectional stack of forward/backward Mamba2 blocks. A single state-space recurrence maps an input $x_t$ to output $y_t$ through a latent state $h_t$:
\begin{equation}
  h_t = \overline{\mathbf{A}}\,h_{t-1} + \overline{\mathbf{B}}\,x_t,
  \qquad
  y_t = \mathbf{C}\,h_t,
  \label{eq:ssm}
\end{equation}
where the continuous parameters $(\mathbf{A},\mathbf{B})$ are discretised with a learnable timestep $\Delta$ via a zero-order hold, $\overline{\mathbf{A}} = \exp(\Delta\mathbf{A})$. This yields linear-time scaling in sequence length, in contrast to the quadratic cost of self-attention. EEGMamba is pre-trained by reconstructing masked patches with a mean-squared-error objective. We load the public checkpoint and fine-tune the full model.

\subsection{Preprocessing}
\label{sec:preproc}

Both datasets are processed with MNE-Python~\cite{gramfort2013} and MNE-BIDS~\cite{appelhoff2019}, with dataset-specific steps reflecting their different recording conditions. 

For \textbf{UGR-MINDVOICE}, we apply a zero-phase FIR band-pass filter (0.1--75\,Hz), retaining the broadband gamma range linked to speech encoding~\cite{lopez2022}, and a 50\,Hz notch for power-line interference. Ocular artefacts are removed with ICA, using EOG references for component identification; the reference channels are then discarded, retaining the 62 scalp channels. Signals are re-referenced to the common average and resampled to 200\,Hz. Stimulus-locked epochs (1.5\,s) are extracted around the speech-production cue and baseline-corrected on the pre-stimulus interval. Subject 15 was excluded from further analysis due to ICA preprocessing errors.

For \textbf{BCIC2020-T3}, the released data are already epoched into short (2\,s) imagined-speech segments, on which a 0.1\,Hz high-pass is not feasible; we therefore apply an IIR band-pass (0.5--75\,Hz) and notches at both 50 and 60\,Hz (the recordings use 60\,Hz mains). ICA is not applied, as the public release provides pre-segmented epochs without EOG channels. Signals are re-referenced to the common average and resampled to 200\,Hz. We use the 2\,s imagined-speech window; using the full 3\,s available segment yielded equivalent results.

Normalisation is applied per model on both datasets: the convolutional baselines receive per-channel $z$-scored input, whereas the foundation models receive physical $\mu$V scaled by $1/100$, matching the scale used during their pre-training.

\subsection{Experimental protocol}
\label{sec:setup}

We evaluate all models using leave-one-subject-out (LOSO) cross-validation.
The dataset contains 14 subjects. In each fold, one subject is set aside as
the test subject and the model is trained on the remaining 13 subjects.
The model is never updated using the test subject's data; the test subject
is only used to decide when to stop training early (after 10 consecutive
epochs without improvement) and to compute the final accuracy for that fold.
This is repeated 14 times, once for each subject, giving 14 independent
test scores. We report the mean and standard deviation across those 14 scores.
For the semantic-category and word tasks, overt and covert recordings of
the same word are treated as the same label, since we are interested in
\emph{what} was said rather than \emph{how} it was said. The
overt-vs-covert distinction is handled separately by the speech-mode task.

We train every model end-to-end with the Adam optimiser~\cite{kingma2015} ($\eta = 10^{-3}$, weight decay $10^{-4}$), batch size 128, cross-entropy loss, for up to 100 epochs with early stopping (patience 10) on the held-out-subject validation accuracy; the best checkpoint is restored for evaluation. The same protocol is used for all models: the baselines (EEGNet, ShallowFBCSPNet, Deep4Net, EEGConformer) are trained from random initialisation, whereas the foundation models are fine-tuned from their public pre-trained checkpoints. 
On BCIC2020-T3, the baselines use $\eta=10^{-3}$ and the foundation models $\eta=10^{-4}$, as $10^{-3}$ destabilises the pre-trained weights.

For BCIC2020-T3 we report two evaluation protocols. First, a \emph{within-subject} protocol with per-subject train/validation splits, as the dataset mandates subject-dependent analysis. Second, a \emph{leave-one-subject-out} (LOSO) protocol.

\subsection{Metrics}
\label{sec:metrics}

Following LaBraM and EEGMamba, we report Balanced Accuracy, Cohen's Kappa, and Weighted F1 for multi-class tasks, as mean$\pm$SD across splits/seeds. We test significance against chance with a Wilcoxon signed-rank test, and compare models pairwise with the same test.

\section{Results}
\label{sec:results}
Table~\ref{tab:main} reports the main benchmark on UGR-MINDVOICE and BCIC2020-T3 datasets across the three tasks of increasing linguistic difficulty, under the LOSO protocol (mean$\pm$SD over 14 subjects).

\begin{table*}[htbp]
  \caption{Main benchmark on UGR-MINDVOICE: foundation models vs.\ convolutional
    baselines. Acc = accuracy (\%) with 95\% CI over 14 LOSO folds (t-distribution),
    W-F1 = weighted F1, $\kappa$ = Cohen's kappa. Best per column in bold.
    Chance accuracy is 1/classes ($\kappa = 0$).
    $^\dagger$EEGConformer collapsed to the majority class on every fold (zero variance).}
  \label{tab:main}
  \centering
  \small
  \setlength{\tabcolsep}{3pt}
  \begin{tabular}{l c ccc ccc ccc}
    \toprule
    & & \multicolumn{3}{c}{\textbf{Speech Mode (3-cl)}}
      & \multicolumn{3}{c}{\textbf{Semantic Category (6-cl)}}
      & \multicolumn{3}{c}{\textbf{Word (60-cl)}} \\
    \cmidrule(lr){3-5}\cmidrule(lr){6-8}\cmidrule(lr){9-11}
    \textbf{Model} & \textbf{Params}
      & \textbf{Acc [95\%~CI]} & \textbf{W-F1} & \textbf{$\kappa$}
      & \textbf{Acc [95\%~CI]} & \textbf{W-F1} & \textbf{$\kappa$}
      & \textbf{Acc [95\%~CI]} & \textbf{W-F1} & \textbf{$\kappa$} \\
    \midrule
    \multicolumn{11}{l}{\emph{Baselines (CNN / hybrid)}} \\
    EEGNet
      & 16.6K
      & \textbf{61.3 [56.5, 66.1]} & \textbf{0.607} & \textbf{0.419}
      & \textbf{19.9 [18.9, 20.8]} & \textbf{0.186} & \textbf{0.039}
      & 2.8 [2.5, 3.1]            & 0.019         & 0.012 \\
    ShallowFBCSPNet
      & 161.6K
      & 33.9 [33.3, 34.4] & 0.207 & 0.008
      & 17.7 [17.2, 18.3] & 0.093 & 0.013
      & 2.1 [1.9, 2.2]    & 0.004 & 0.004 \\
    Deep4Net
      & 0.3M
      & 34.4 [33.7, 35.0] & 0.230 & 0.015
      & 17.4 [17.1, 17.8] & 0.073 & 0.009
      & 2.0 [1.8, 2.2]    & 0.002 & 0.004 \\
    EEGConformer$^\dagger$
      & 405.9K
      & 33.3\phantom{[00.0, 00.0]} & 0.167 & 0.000
      & 16.7\phantom{[00.0, 00.0]} & 0.048 & 0.000
      & 1.7\phantom{[0.0, 0.0]}   & 0.001 & 0.000 \\
    \midrule
    \multicolumn{11}{l}{\emph{Foundation models (pre-trained, fine-tuned)}} \\
    LaBraM-Base
      & 5.8M
      & 57.3 [53.0, 61.7] & 0.562 & 0.360
      & 17.0 [16.6, 17.3] & 0.056 & 0.003
      & 1.9 [1.7, 2.1]    & 0.002 & 0.003 \\
    EEGMamba
      & 8.3M
      & 56.6 [51.8, 61.3] & 0.553 & 0.349
      & 19.3 [18.5, 20.1] & 0.150 & 0.032
      & \textbf{2.9 [2.5, 3.3]} & \textbf{0.023} & \textbf{0.012} \\
    \midrule
    Chance & -- & 33.3 & -- & 0 & 16.7 & -- & 0 & 1.7 & -- & 0 \\
    \bottomrule
  \end{tabular}
\end{table*}

\begin{table*}[t]
  \caption{BCIC2020-T3 imagined speech (5 classes), within-subject and leave-one-subject-out (LOSO). Bal.Acc = balanced accuracy (\%), W-F1 = weighted F1, $\kappa$ = Cohen's kappa; mean$\pm$SD over 15 subjects. Chance accuracy $= 20\%$ ($\kappa=0$). Best within-subject value per column in bold. Under LOSO no model exceeds chance (all $p>0.18$).}
  \label{tab:bcic}
  \centering
  \small
  \setlength{\tabcolsep}{4pt}
  \begin{tabular}{l c ccc ccc}
    \toprule
    & & \multicolumn{3}{c}{\textbf{Within-subject}} & \multicolumn{3}{c}{\textbf{LOSO (cross-subject)}} \\
    \cmidrule(lr){3-5}\cmidrule(lr){6-8}
    \textbf{Model} & \textbf{Params} & \textbf{Bal.Acc} & \textbf{$\kappa$} & \textbf{W-F1} & \textbf{Bal.Acc} & \textbf{$\kappa$} & \textbf{W-F1} \\
    \midrule
    \multicolumn{8}{l}{\emph{Baselines (CNN / hybrid)}} \\
    EEGNet          & 3.1K   & $26.7{\pm}3.3$ & $0.084{\pm}0.041$ & $0.245{\pm}0.033$ & $18.9{\pm}1.6$ & $-0.014{\pm}0.020$ & $0.184{\pm}0.017$ \\
    ShallowFBCSPNet & 107.7K & $28.7{\pm}4.4$ & $0.109{\pm}0.055$ & $0.250{\pm}0.058$ & $19.4{\pm}2.7$ & $-0.007{\pm}0.034$ & $0.183{\pm}0.027$ \\
    EEGConformer    & 447.2K & $\mathbf{30.2{\pm}4.4}$ & $\mathbf{0.127{\pm}0.055}$ & $\mathbf{0.258{\pm}0.052}$ & $19.5{\pm}1.5$ & $-0.006{\pm}0.018$ & $0.146{\pm}0.029$ \\
    \midrule
    \multicolumn{8}{l}{\emph{Foundation models}} \\
    LaBraM-Base     & 5.8M & $27.7{\pm}2.7$ & $0.096{\pm}0.034$ & $0.228{\pm}0.038$ & $20.2{\pm}1.2$ & $\phantom{-}0.002{\pm}0.014$ & $0.109{\pm}0.038$ \\
    EEGMamba        & 3.4M & $25.7{\pm}2.5$ & $0.071{\pm}0.031$ & $0.180{\pm}0.034$ & $19.3{\pm}1.9$ & $-0.009{\pm}0.024$ & $0.139{\pm}0.027$ \\
    \midrule
    Chance          & --   & 20.0 & 0.00 & -- & 20.0 & 0.00 & -- \\
    \bottomrule
  \end{tabular}
\end{table*}
\subsection{Foundation models vs.\ baselines}
\begin{figure}[t]
      \centering
      \includegraphics[width=\columnwidth]{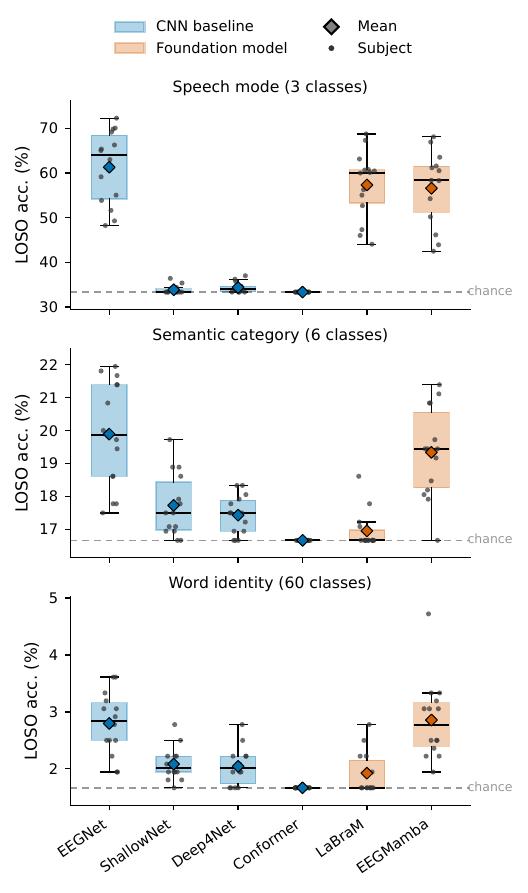}
      \caption{Per-subject leave-one-subject-out accuracy on UGR-MINDVOICE for the three tasks of increasing linguistic difficulty (stacked top to bottom). Each box shows the distribution over the 14 held-out subjects; dots are individual subjects, the diamond marks the mean, and the dashed line marks chance ($1/\text{classes}$).}
      \label{fig:per_subject}
    \end{figure}

    On UGRMINDVOICE the central observation is that neither pre-trained foundation model outperforms the compact EEGNet. On speech-mode decoding (overt/covert/rest, chance \SI{33.3}{\percent}), EEGNet attains \SI{61.3}{\percent}, exceeding both LaBraM (\SI{57.3}{\percent}) and EEGMamba (\SI{56.6}{\percent}) by 4--5 points and far surpassing the remaining baselines, which barely leave chance (ShallowFBCSPNet \SI{33.9}{\percent}, Deep4Net \SI{34.4}{\percent}, EEGConformer \SI{33.3}{\percent}). On semantic-category decoding (chance \SI{16.7}{\percent}) EEGNet leads at \SI{19.9}{\percent}, with EEGMamba close behind (\SI{19.3}{\percent}) and LaBraM falling back to chance (\SI{17.0}{\percent}); on word identity (chance \SI{1.7}{\percent}) every model sits within roughly one point of chance, EEGMamba (\SI{2.9}{\percent}) and EEGNet (\SI{2.8}{\percent}) nominally highest and LaBraM (\SI{1.9}{\percent}) at chance. Thus the foundation models either trail EEGNet or, on the harder lexical tasks, collapse to chance, despite having roughly $350$--$500\times$ more parameters; EEGMamba's nominal edge on the word task amounts to a single additional point above chance and is not meaningful. Cohen's $\kappa$ tells the same story: only EEGNet ($\kappa=0.42$), LaBraM ($\kappa=0.36$) and EEGMamba ($\kappa=0.35$) show substantial above-chance agreement, all on speech mode, whereas on the category and word tasks every model has $\kappa<0.04$ (Table~\ref{tab:main}), confirming that the near-chance accuracies reflect genuinely uninformative predictions rather than a class-imbalance artefact. A Wilcoxon signed-rank test over the 14 per-subject scores sharpens this picture. Every model except EEGConformer decodes significantly above chance on all three tasks (one-sided, $p<0.05$) --- including the lexical tasks, where the above-chance margins are nonetheless tiny --- while EEGConformer, pinned exactly at chance, is the sole exception ($p=1.0$). Also, no model significantly \emph{exceeds} EEGNet on any task (paired two-sided test): on speech mode EEGNet significantly outperforms both LaBraM ($\Delta=-4.0$ points, $p=0.007$) and EEGMamba ($\Delta=-4.7$, $p<0.001$), and on the lexical tasks the foundation models at best match it --- EEGMamba is statistically indistinguishable from EEGNet on semantic category ($p=0.25$) and word identity ($p=0.79$), while LaBraM remains significantly below it ($p\le0.013$). The per-subject distributions in Fig.~\ref{fig:per_subject} show that this pattern is consistent across the cohort rather than driven by a few outliers. This stands in sharp contrast to the consistent gains foundation models report on motor imagery, seizure and sleep tasks. We further note that EEGConformer collapses to the majority/constant prediction on every task (accuracy exactly at chance, zero variance), indicating it fails to train in this low-SNR, subject-independent regime; we report it unmodified. 

On BCIC2020-T3, all five models decoded imagined speech significantly above the 20\% chance level (one-sided Wilcoxon, $p = 0.0003$), but balanced accuracy remained low (25.7--30.2\%, Table~\ref{tab:bcic}). Neither foundation model outperformed the convolutional baselines: LaBraM did not significantly exceed any baseline (vs.\ EEGNet, $+1.0$\,pp, $p = 0.41$), and the highest balanced accuracy was obtained by a baseline (EEGConformer, 30.2\%). EEGMamba performed below the stronger baselines (vs.\ EEGConformer, $-4.5$\,pp, $p < 0.001$). Despite $\sim$16{,}724\,h (EEGMamba) and $\sim$2{,}500\,h (LaBraM) of pre-training and roughly $1000\times$ more parameters, large-scale EEG pre-training conferred no advantage over a 3K-parameter CNN; this held under both a 2\,s imagery and a full 3\,s window.

Under the leave-one-subject-out protocol on BCIC2020-T3, all five models performed at the 20\% chance level (balanced accuracy 0.189--0.202; Cohen's $\kappa \approx 0$), and none exceeded chance (one-sided Wilcoxon, all $p > 0.18$). No foundation model meaningfully outperformed the baselines (all $|\Delta| \leq 0.013$). Cross-subject imagined-speech decoding thus remains unsolved on this dataset, and large-scale pre-training did not enable transfer to unseen subjects any more than a compact CNN did.

\subsection{Overt vs.\ Covert }
Decoding degrades steeply as the target shifts from speech mode to lexical content. The mode task --- which only requires distinguishing overt, covert and rest states --- is decoded well above chance, but semantic-category and word-level decoding fall to near chance for all six models. To test whether this lexical-decoding failure is specific to imagined (covert) speech, we re-evaluated each pooled-trained LOSO model separately on the held-out subject's overt-only and covert-only epochs. The two speaking modes are statistically indistinguishable for every model on both lexical tasks (paired Wilcoxon signed-rank over the 14 subjects, all $p>0.11$): on semantic category the overt/covert accuracies are, for example, \SI{18.8}{\percent}/\SI{21.0}{\percent} (EEGNet) and \SI{18.9}{\percent}/\SI{19.8}{\percent} (EEGMamba), and on word identity overt and covert agree to within \SI{0.2}{\percent} for every model --- in both cases near chance. Lexical content is therefore no more decodable from overt than from covert speech in this corpus; the only robustly decodable signal is the overt/covert/rest distinction itself.

\section{Conclusion}
\label{sec:conclusion}

We presented the first systematic benchmark of EEG foundation models against convolutional baselines for overt, covert and imagined speech decoding, on two complementary corpora including a new leakage-free Iberian-Spanish dataset. Across both corpora the finding is consistent: neither LaBraM nor EEGMamba outperformed a compact EEGNet. On UGR-MINDVOICE, EEGNet gave the best speech-mode decoding (\SI{61.3}{\percent} vs.\ LaBraM's \SI{57.3}{\percent} and EEGMamba's \SI{56.6}{\percent}), while semantic-category and word decoding stayed near chance for every model; on BCIC2020-T3 imagined speech, the highest balanced accuracy came from a baseline (EEGConformer, \SI{30.2}{\percent}), no foundation model exceeded the baselines, and all models fell to chance under cross-subject evaluation. Our results therefore indicate that current general-purpose EEG pre-training on motor imagery, sleep and seizure data do not transfer reliably to the problem of speech and language decoding and subject-adaptive fine-tuning as priorities for closing the gap with invasive systems. Future work will add multi-seed runs to complement the across-subject significance tests reported here and explore speech-specific pre-training.

\section{Acknowledgements}
\ifcameraready
    This work was supported by grants PID2022-141378OB-C22 and AIA2025-163317-C32 funded by MICIU/AEI/10.13039/501100011033 and ERDF/EU.
\else
    Acknowledgements withheld for double-blind review.
\fi

\section{Generative AI Use Disclosure}
Claude (Anthropic) was used to assist with manuscript structuring and editing. The authors reviewed and edited the content and take full responsibility for the final version.

\bibliographystyle{IEEEtran}
\bibliography{mybib}

\end{document}